# The impact of composition choices on solar evolution: age, helio- and asteroseismology, and neutrinos

Diogo Capelo  ⋆ and Ilídio Lopes
*Centro de Astrofísica e Gravitação – CENTRA, Departamento de Física, Instituto Superior Técnico – IST, Universidade de Lisboa – UL, Av. Rovisco Pais 1, P-1049-001 Lisboa, Portugal*

**ABSTRACT**
The Sun is the most studied and well-known star, and as such, solar fundamental parameters are often used to bridge gaps in the knowledge of other stars, when these are required for modelling. However, the two most powerful and precise independent methodologies currently available to infer the internal solar structure are in disagreement. We aim to show the potential impact of composition choices in the overall evolution of a star, using the Sun as example. To this effect, we create two Standard Solar Models and a comparison model using different combinations of metallicity and relative element abundances and compare evolutionary, helioseismic, and neutrino-related properties for each. We report differences in age for models calibrated to the same point on the HR diagram, in red giant branch, of more than 1 Gyr, and found that the current precision level of asteroseismic measurements is enough to differentiate these models, which would exhibit differences in period spacing of 1.30–2.58 per cent. Additionally, we show that the measurement of neutrino fluxes from the carbon–nitrogen–oxygen cycle with a precision of around 17 per cent, which could be achieved by the next generation of solar neutrino experiments, could help resolve the stellar abundance problem.

## 1 INTRODUCTION

The interior of stars, for many years thought unobservable, has been probed for the past 40–50 yr thanks to the efforts of helio- and asteroseismology (Leighton, Noyes & Simon 1962; Ulrich 1970), even if inversions for the solar and stellar cores are still limited, due to insufficient data. Likewise, in the last half century, the study of neutrinos has motivated a string of scientific successes, whose examples include the prediction of the Nobel prize-winning solar neutrino fluxes (Ahmad et al. 2001), instrumental in settling the Solar Neutrino Problem (SNP) first encountered by Davis, Harmer & Hoffman (1968) and Bahcall & Shaviv (1968) and confirming the theory of neutrino oscillations. Interplay between helioseismology and solar neutrinos has always occurred, with the former confirming solar models and motivating a search for the solution to the SNP in the domain of the behaviour of the particles rather than a revision of stellar physics. Since then, solar neutrino experiments themselves have more recently began to emerge as a potential probe of regions traditionally inaccessible to helioseismology (Lopes & Turck-Chièze 2013). Neutrinos are also hypothesized to be relevant to the determination of the density of the solar core (Lopes 2013), something that will be especially relevant when measurements sensitive enough to detect matter oscillations start to be obtained for more sources, and something helioseismic techniques are sensitive to.

Still, challenges remain to our understanding of the Sun, especially its composition, which is – alongside the mass – one of the main factors driving stellar evolution. Despite the extraordinary advances in spectroscopic analysis techniques since circa the year 2000, when researchers (Asplund, Grevesse & Sauval 2005; Grevesse, Asplund & Sauval 2007; Scott et al. 2015a) began developing and using three-dimensional radiation hydrodynamical models of the solar atmosphere and improved meteorite analysis (Lodders 2003; Lodders, Palme & Gail 2009) to complement observations of the solar surface, several works (Guzik & Watson 2004; Haxton & Serenelli 2008; Ramirez, Melendez & Asplund 2009) indicate that determinations made exclusively in the surface may not be a good indicator of the star's overall metallicity, and particularly, stars exhibiting the same surface metallicities may have different internal structures resulting in different distributions of the relative abundances of elements in their interiors.

The so-called abundance problem arises from apparent contradictions in the predictions for the internal structure of the Sun made by solar models using the more sophisticated metallicity assumptions and high-precision results for the speed of propagation of acoustic waves inside the Sun (Bahcall, Pinsonneault & Wasserburg 1995; Gough et al. 1996; Christensen-Dalsgaard 2002). The improvement of spectroscopic data analysis confirmed and continued an ongoing trend of substantial reductions to the inferred abundances of, in particular, the carbon–nitrogen–oxygen (CNO) cycle's elements and the revision of the overall ratio of metals to hydrogen at the surface, $(Z/X)_s$, which was changed from the pre-solar atmosphere modelling value of 0.0231 in Grevesse & Sauval (1998) (GS98) to 0.0181 in Asplund et al. (2009) (AGSS09). This decrease in the inferred metallicity of the Sun shifted Standard Solar Models' (SSM) predictions of its internal structure, whereas the continued improvements to helioseismic techniques and data increased the precision of the previous results, but did not substantially alter the quality of the determinations regarding internal structure, destroying the previous relative agreement between spectroscopy and helioseismology (Basu et al. 2009; Turck-Chièze, Piau & Couvidat 2011). Furthermore, advancements in helioseismology saw the extension of the field to a general study of asteroseismology, with the advent of the *Kepler* mission detecting more than 500 solar-type oscillators (Chaplin et al. 2011), and around 14 000 pulsating red giants, which seemed to confirm, rather than prompt a revision of, the solar results.

⋆ E-mail: diogo.capelo@tecnico.ulisboa.pt

The uncertainty resulting from the differing predictions of spectroscopy and helioseismology poses problems to the determination of general stellar properties (namely total mass, radius, luminosity, effective temperature, metallicity, and element abundances) for many stars, including the *Kepler* mission stars, many of which lack precise measurements in those fundamental quantities. This is because estimation of those stellar quantities has had to rely upon the use of global spectroscopic or photometric observation of surface properties (colours and surface metallicity) that, as will be shown for the case of the Sun, may be insufficient to adequately inform, or even extrapolate, the internal characteristics of the star. Asteroseismic analysis of individual oscillation frequencies is instrumental in discriminating stars with similar surface quantities and different structures (Chaplin & Miglio 2013), but this is not sufficient because asteroseismic data themselves require prior constraints on quantities like metallicity to estimate parameters like mass and age.

Similarly to how the overlap between asteroseismology and spectroscopy permitted researchers to differentiate the interiors of different stars with similar surfaces, we aim to study whether or not solar neutrinos can be used in conjunction with other established techniques as a way to differentiate the cores or innermost regions of the star, where those other techniques are difficult to apply, by obtaining several metallicity models and comparing the predicted results of each in various stages of evolution.

In this work, we found that, as proposed by Cerdeño et al. (2018), the measurement of CNO solar neutrino fluxes is possible (Lopes, Silk & Hansen 2002) and necessary if we wish to correctly predict the correct amount of metals inside the Sun and stars, and in doing so be able to make the correct inference about the structure of Sun-like interiors in more advanced stages of stellar evolution, as it will very likely be possible with data from the forthcoming *PLATO* mission.

The following work is divided into five main sections: In Section 2, we explain how our SSM is built and how it compares with other similar models in the literature; in Section 3, we study and compare surface properties between the models; in Section 4, we focus on helioseismic-motivated properties; in Section 5, we analyse neutrino emission both as a whole and per source while comparing the results to those obtained in the previous sections; and finally in Section 6 we summarize and present our results.

## 2 MODEL BUILDING

The first step in studying the impact of chemical composition in solar evolution is to build an SSM, which will then be allowed to continue evolving to provide a picture of its entire evolution until the red giant phase. In general terms, an SSM is a specific class of one-dimensional stellar evolution model that has been generated for a mass of 1 $M_\odot$ and allowed to evolve in time until the present-day solar age, $\tau_\odot = 4.57$ Gyr (Bahcall, Pinsonneault & Wasserburg 1995; Bahcall, Serenelli & Basu 2006).

We obtain the solar model of the present Sun using a $\chi^2$ calibration method to ensure that our SSM fits all the observational constraints.

**Table 1.** Logarithmic abundances $\log \epsilon_i \equiv \log N_i/N_H + 12$ of the most relevant metals in solar modelling for the two considered high- (GS98) and low-metallicity (AGSS09) determinations. $N_i$ is the number abundance of the element $i$.

| Element | GS98 | AGSS09 |
| --- | --- | --- |
| C | 8.52 | 8.43 |
| N | 7.92 | 7.83 |
| O | 8.83 | 8.69 |
| Ne | 8.08 | 7.93 |
| Mg | 7.58 | 7.60 |
| Si | 7.55 | 7.51 |
| S | 7.33 | 7.13 |
| Ar | 6.40 | 6.40 |
| Fe | 7.50 | 7.50 |
| $(Z_s/X_s)_\odot$ | 0.0231 | 0.0181 |

This procedure is repeated several times until we obtain a stellar model with a minimum total $\chi^2$, composed of the individual contributions of each of the considered observational targets. The free parameters of the $\chi^2$ calibration procedure are the following ones: initial helium mass fraction ($Y_i$), initial metal mass fraction ($Z_i$), mixing length parameter ($\alpha_{MLT}$), and overshoot parameter ($f_{ov}$). The solar observable quantities used in the $\chi^2$ for the calibration are luminosity and effective temperature, respectively, $L_\odot = 3.8418 \times 10^{33}$ erg s$^{-1}$ and $T_{eff} = 5777$ K, the helium mass fraction at the surface ($Y_s = 0.2485$), and the metal-to-hydrogen abundance ratio at the surface, ($Z_s/X_s$). This last ratio depends on the element abundances adopted for a given stellar model. Table 1 shows the ($Z_s/X_s$) values used in our models, and more information about the solar observational data is available in Bahcall et al. (2005) and Turck-Chièze & Couvidat (2011). Additionally, the High-$Z$ model is also calibrated to match the helioseismic determination of the location of the base of the convective zone, $R_{CZ}/R_\odot = 0.713$, from Basu & Antia (2004), and the sound speed profile in the interior of the solar medium, $c_\odot$, from Basu et al. (2009). For each model, the weights of all terms allowed to contribute to the $\chi^2$ are equal.

### 2.1 Input physics

The release version 12115 of the stellar evolution code MESA (Paxton et al. 2011, 2013, 2015, 2018, 2019) was used to generate the models from the pre-main sequence (PMS), assuming that the Sun was initially chemically homogeneous and fully convective. Moreover, the evolution code used to obtain the models was modified to allow the direct computation of radial profiles of neutrino emission from the nuclear reactions pertaining to the pp chain and CNO cycle, e.g. Lopes & Turck-Chièze (2013). The input physics considers diffusion of helium and metals (including gravitational settling) using the method of Thoul, Bahcall & Loeb (1994) and convection follows the mixing length theory (Böhm-Vitense 1958). To account for the hydrodynamical mixing instabilities at the convective boundary, a parametric model of convective overshoot is used (Paxton et al. 2011). The atmosphere is a grey Eddington model. Nuclear reactions rates were taken from the (updated) JINA REACLIB (Cyburt et al. 2010) – with the exception of the $^9$Be($\alpha$,n)$^{12}$C reaction, which is taken from Kunz et al. (1996) – under weak (Salpeter) screening (Salpeter 1954), which provides a more fitting description of the solar process (Gruzinov & Bahcall 1998) than the other already implemented methods.

## 2.2 Chemical composition

Two composition tables were selected from among the several available in the literature for their representativity of different classes of abundance determinations: the high-metallicity compositions that provided models in better agreement with precision helioseismic data, and the more recent low-metallicity compositions, that traditionally do not agree as well with observation, but that have been obtained using considerably more refined methods.

Given their level of resilience and exposure in the scientific community, the two composition determinations chosen to represent, respectively, the high- and low-metallicity classes of solar abundances for this study were the ones presented in GS98 (High-Z) and AGSS09 (Low-Z). The distribution of some key elements in each of these solar mixtures can be found in Table 1.

## 2.3 Equation of state and opacities

The choice of composition is linked with the calculation of opacities and therefore with the equation of state (EoS), since contributions to opacity from atomic transitions will vary based on what elements are available in the star and in what quantities, among other processes like ionization and excitation of ions that depend on other stellar properties, like temperature. Opacity will then affect the radiative energy transfer and so, in principle, the composition reflected in the model should be reflected in the opacity tables and the EoS as well. In practice, most recent solar models use the OPAL EoS (Rogers & Nayfonov 2002) or its 2005 update, which accounts only for a mixture of six metals, up to Ne. The MESA code offers versions of the OPAL EoS pre-compiled with abundances according to both GS98 and AGSS09 compositions, which allows consistency to be retained for the models presented in this work, with only the caveat of not referencing metals more massive than Ne, a characteristic that does not significantly affect the validity of the models (Vinyoles et al. 2017).

The atomic opacities used are from the Opacity Project (Badnell et al. 2005), corrected for low temperatures with molecular opacities from Ferguson et al. (2005). It should be noted that opacities are another important factor, and it has been found that a maximum increase of about 22 per cent in opacity at the base of the convective zone (Christensen-Dalsgaard & Houdek 2010) would be enough to reconcile the results of high- and low-metallicity models and bring the latter into agreement with helioseismology (Christensen-Dalsgaard et al. 2009). Additionally, recent pioneering efforts to measure opacities directly in a laboratory environment, reproducing the conditions of the solar interior, have found a stark disagreement between experiment and theory, with the measured wavelength-dependent opacity for Fe, which contributes with a quarter of total opacity at the radiative/convective boundary, being 30–400 per cent higher than that predicted (Bailey et al. 2015). Even though this result accounts for only half of the necessary change in total mean opacity to explain the discrepancy between solar models and helioseismology, it is a clear indication that significant improvements are possible in opacity calculations.

## 2.4 Converged model parameters

A summary of parameters relevant to the generation of the models and to the helioseismology of the current-age Sun can be seen in Table 2 for the high-metallicity GS98 (High-Z) and low-metallicity AGSS09 (Low-Z) compositions. We also compare our results against those of Vinyoles et al. (2017), who have obtained SSMs with similar

**Table 2.** Best values of the main relevant physical quantities extracted from the models and comparison with other SSMs – V-H corresponds to B16-GS98 and V-L to B16-AGSS09met from Vinyoles et al. (2017) – and with observed values, when measurements exist: $R_{CZ}/R_\odot$ from Basu & Antia (2004), $Y_s$ from Basu & Antia (1997), and $c_\odot$ for the calculation of $\delta c/c$ from Basu et al. (2009).

| Quantity | High-Z | V-H | Low-Z | V-L | Sun |
|---|---|---|---|---|---|
| $Y_s$ | 0.2472 | 0.2426 | 0.2352 | 0.2317 | $0.2485 \pm 0.0035$ |
| $R_{CZ}/R_\odot$ | 0.7136 | 0.7116 | 0.7326 | 0.7223 | $0.713 \pm 0.001$ |
| $\langle \delta c/c \rangle$ | 0.0007 | 0.0005 | 0.0062 | 0.0021 | 0 |
| $\alpha_{MLT}$ | 1.90 | 2.18 | 1.75 | 2.11 | – |
| $Y_i$ | 0.2719 | 0.2718 | 0.2609 | 0.2613 | – |
| $Z_i$ | 0.0187 | 0.0187 | 0.0160 | 0.0149 | – |
| $(Z/X)_s$ | 0.0235 | 0.0230 | 0.0181 | 0.0178 | – |

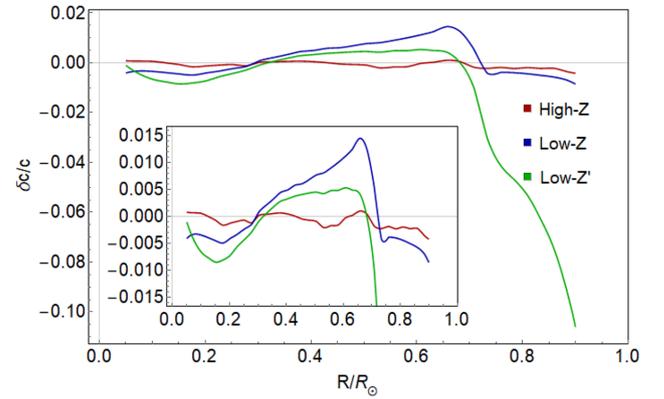

**Figure 1.** Relative sound speed difference during the main sequence between observation and models using the GS98 (High-Z, in red) and AGSS09 (Low-Z, in blue). Also plotted is a model calibrated to match point RG2, using a GS98 composition but a low metallicity (Low-Z', in green). The solar reference value is taken from Basu et al. (2009).

inputs to ours and have done a very complete and detailed study of their models.

The relative sound speed difference between the reference and the model, $\delta c/c$, is defined as

$$\frac{\delta c}{c} = \frac{c_\odot - c_{model}}{c_\odot}, \quad (1)$$

where $c_\odot$ is the sound speed for the solar observational reference (Basu et al. 2009) and $c_{model}$ is the solar sound speed for the considered model. Accordingly, we calculate the ratio $\delta c/c$ (equation 1) for several radial distances $r_i$, a discrete set of values of the stellar radius, $i = 1, ..., N$ such that $r = 0$ ($i = 1$) and $R_\odot$ ($i = N$), allowing us to compute the following average of the $\delta c/c$ ratios:

$$\left\langle \frac{\delta c}{c} \right\rangle = \frac{1}{N} \sum_{i=0}^{N} \sqrt{\left(\frac{\delta c}{c}\right)_{r_i}^2}. \quad (2)$$

This quantity gives us a measure of the overall averaged discrepancy of the speed of sound speed profile of a given solar model and the helioseismic sound speed. For the present case, $N = 37$, corresponding to 37 radial data points. A plot of the relative sound speed difference for all models can be found in Fig. 1.

The main disagreement between our models and those of Vinyoles et al. (2017) is in the parameter $\alpha_{MLT}$. This can be explained by the difference in considered atmosphere modelling. While we consider a simple Eddington grey atmosphere, Vinyoles et al. (2017) consider a Krishna–Swamy atmosphere. Since our work is concerned mostly with what occurs in the central regions of the star, the difference in considered atmospheric models should not have a direct impact in the work, and the indirect impact it could have through its effect in the model's converged parameters is not relevant, since these parameters are not, in any case, significantly different from the ones in the literature.

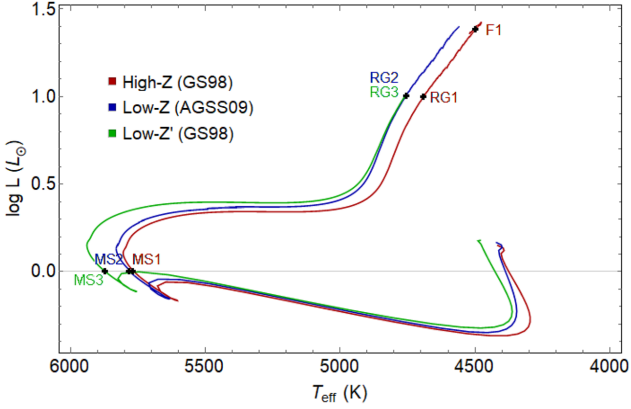

**Figure 2.** Hertzsprung–Russel diagram for models of a 1 $M_\odot$ star using the standard High-$Z$ (GS98, in red) and Low-$Z$ (AGSS09, in blue) compositions. Also plotted (in green) is a model calibrated to match point RG2 from the Low-$Z$ model track. This model is named Low-$Z'$ since its metallicity is close to the Low-$Z$ model's one, but the individual relative abundances of elements follow the distribution of GS98.

The results of the models fall within the expectation established by the many other published works so far, namely, for the High-$Z$ model, all relevant solar quantities are obtained within experimental uncertainties and the common diagnostic tools from helioseismology, the surface helium abundance and the position of the base of the convective zone, agree well with the obtained results. Although most authors use distinct calibration procedures to compute solar models, leading to predictions not directly comparable, we present here, for reference, some of these results. Our differences between model and observation wield $\chi^2 = 1.04$ for the High-$Z$, which is comparable to 0.91 from Vinyoles et al. (2017), the main source of error being the solar speed profile matching. For the Low-$Z$ model, the discrepancies become more significant, failing to match the experimental targets as well, evidence of the solar abundance problem. For this model, we obtain a $\chi^2 = 7.94$ that is higher than that of 6.45 for Vinyoles et al. (2017), achieving the target values for the helium surface fraction and surface relative metallicity, $(Z/X)_s$.

Comparison of our High-$Z$ model with the MESA solar model of Paxton et al. (2011) (MESA-I), obtained using the same composition and a similar calibration procedure, reveals slight improvements in reproducing the surface helium fraction $Y_s$, whose value in MESA-I is 0.2433, matching the observed value at $2\sigma$, whereas the $Y_s$ value of High-$Z$ (0.2472) matches observation in a $1\sigma$ interval. All other quantities are similar between both models, which is to be expected since both take advantage of the same capabilities of MESA.

## 3 SOLAR SURFACE AND EVOLUTION

After calibrating the models for the current solar age, they were allowed to evolve further, until the stars reached the red giant phase, in order to highlight the impact of the different compositions in their later lives. The evolutionary tracks for the two models from the PMS up to the red giant branch can be seen in Fig. 2. The effect of the calibration can be seen by the overlap of $L$ and $T_{eff}$ at the current age in points MS1 and MS2.

**Table 3.** Best values of the initial parameters determined for the comparison model Low-$Z'$. The same quantities for the two SSMs High-$Z$ and Low-$Z$ are also presented here again.

| Quantity | Low-$Z'$ | High-$Z$ | Low-$Z$ |
|---|---|---|---|
| $\alpha_{MLT}$ | 1.88 | 1.90 | 1.75 |
| $f_{ov}$ | 0.018 | 0.019 | 0.020 |
| $Y_i$ | 0.2739 | 0.2719 | 0.2609 |
| $Z_i$ | 0.0165 | 0.0187 | 0.0160 |

The rapid changes that characterize post-MS stellar evolution mean that, when left to evolve separately, the High-$Z$ and Low-$Z$ models drift significantly apart for the same elapsed times, and comparisons of points of similar age result in large differences, as can be seen in Table 4, by comparing points RG2 (Low-$Z$ model) and F1 (High-$Z$ model). Thusly, in order to compare the effects of composition in the later stages of the evolution of a star, age becomes a somewhat misleading benchmark. In order to have another reasonable candidate for comparison, additionally to the two models already discussed in Section 2, which were calibrated to match current solar observations, a third model – Low-$Z'$ – was produced, using the same relative abundance of elements as the High-$Z$ model (GS98), but calibrated instead to match a specific point (RG2) on the evolutionary track of the Low-$Z$ model. This calibration procedure varies the same initial free parameters as the one for the High-$Z$ and Low-$Z$ models: initial helium ($Y_i$), initial metal mass fraction ($Z_i$), mixing length parameter ($\alpha_{MLT}$), and overshoot parameter ($f_{ov}$), and the same constraints – with the exception of the matching of the location of the base of the convective zone and sound speed profile – though these constraints are now theoretical, as they pertain to the red giant phase of the Low-$Z$ model (point RG2), and not observational, as they were in the case of the calibration of the SSMs. The new Low-$Z'$ model aims to try and measure what impact in the remaining solar parameters changing the relative abundance of elements will have, while still constraining the model to match a point of the evolutionary track obtained with the other class of relative abundance determinations. The converged parameters for this new model can be consulted in Table 3.

This requirement resulted in a model with a metallicity of $Z_{Low-Z'} = 0.0165$, almost identical to the one obtained for Low-$Z$, $Z_{Low-Z} = 0.0160$, hence the terminology Low-$Z'$. In Table 3, it is worth noticing that the models Low-$Z$ and Low-$Z'$, which are located on the same point on the red giant path (RG2 and RG3 in Fig. 2), have very different $\alpha_{MLT}$ and $Y_i$ between them (their difference exceeding that between the High-$Z$ and Low-$Z$ models), indicating differences in the scale of the convection processes in the star, which could help explain its different internal structure. It bears reiterating that the Low-$Z'$ model was constructed to provide a comparison for the other models, and does not intend to be a depiction of the Sun or an SSM.

The main evolutionary parameters pertaining to all three of these models can be found in Table 4.

The High-$Z$ and Low-$Z$ models exhibit very similar surface parameters during the entirety of their stay in the main sequence, and even as they progress into the red giant phase, comparing the models at points of similar luminosity wields a maximum difference of 1.41 per cent in $T_{eff}$ for the models with the same relative element abundances but different metallicities (High-$Z$ and Low-$Z'$). Despite this apparent agreement, a significant discrepancy can be seen in terms of the age, with Low-Z' (RG3) reaching a luminosity comparable to the other two (RG1 and RG2) more than 1 Gyr faster than those models.

**Table 4.** Values of some relevant surface and global quantities for different models, for different stages of evolution.

| Model | | Age (Gyr) | | | $R/R_\odot$ | | | $L/L_\odot$ | | | $T_{\rm eff}$ (K) | | |
|---|---|---|---|---|---|---|---|---|---|---|---|---|---|
| | | Low-Z | High-Z | Low-Z' | Low-Z | High-Z | Low-Z' | Low-Z | High-Z | Low-Z' | Low-Z | High-Z | Low-Z' |
| 1 $M_\odot$ | MS | 4.57 | 4.57 | 2.97 | 0.999 | 0.999 | 0.969 | 0.999 | 1.00 | 1.001 | 5781 | 5772 | 5870 |
| | RG | 12.0 | 11.86 | 10.77 | 4.73 | 4.87 | 4.71 | 10.26 | 10.26 | 10.22 | 4753 | 4686 | 4755 |
| | F1 | – | 12.0 | – | – | 8.42 | – | – | 25.98 | – | – | 4493 | – |

## 4 ASTEROSEISMOLOGY AND THE SOLAR INTERIOR

Helioseismology is the study of oscillations in the solar medium, driven by internal processes (like convection) and with gravity and pressure as restoring forces. A thorough review of helioseismology can be found in Christensen-Dalsgaard (2002).

Since the observation of these oscillations in the Sun, thousands of main-sequence and red-giant solar-like oscillators have been observed by several missions, pioneering the field of asteroseismology, e.g. Aerts, Christensen-Dalsgaard & Kurtz (2010), which is capable of providing information regarding the internal structure of a star by analysing the way these waves propagate inside it.

From the p modes (for which pressure is the restoring force), information can be gathered regarding the envelope of the star, and one commonly used indicator is the large frequency separation, which measures the difference in the frequency between consecutive modes of the same angular degree, $\Delta\nu = \nu_{n,\ell} - \nu_{n-1,\ell}$. This quantity can be shown to relate to the star's mass and radius (Chaplin & Miglio 2013), and depends on the speed of sound in the interior of the star (Tassoul 1980):

$$\Delta\nu = \left(2\int_0^R \frac{{\rm d}r}{c(r)}\right)^{-1}, \qquad (3)$$

where $R$ is the total radius of the star and $c(r)$ is the speed of sound at radius $r$.

On the other hand, g modes probe the inner core where they propagate, and can be described by their separation in period, $\Delta\Pi_\ell$, which relates to the size of this core, e.g. Montalbán et al. (2013), and depends on the Brunt–Väisälä (or buoyancy) frequency (Tassoul 1980):

$$\Delta\Pi_\ell = \frac{2\pi^2}{\sqrt{l(l+1)}}\left(\int_{r_1}^{r_2} N\frac{{\rm d}r}{r}\right)^{-1}, \qquad (4)$$

where $r_1$ and $r_2$ are the turning points of the g-mode cavity, i.e. the limits of the region of the star where the g modes propagate, $\ell$ is the angular degree, and $N$ is the Brunt–Väisälä frequency.

The speed of sound in the stellar medium, which can be obtained by inversion of asteroseismic data, is also a powerful tool for validating SSMs and generally constraining the parameter space of stellar models. As such, it has been determined for each model at the solar age and compared with observation for the current Sun, as presented in Fig. 1.

As stars progress in the red giant branch, the layer where nuclear reactions are occurring in the star changes in radius, which should leave an imprint in the star's properties. Asteroseismology is sensitive to this region, since there is a steep gradient of the hydrogen profile, which is being consumed in a thin shell around the helium core via the pp-chain and CNO-cycle reactions (with around 20–80 per cent relative energy distribution, respectively; see Fig. 5), leading to a noticeable peak in the Brunt–Väisälä frequency (as can be seen in Fig. 4), and the 'bump' in the sound speed profiles (Fig. 3). Correspondingly, a noticeable $\Delta\Pi_1$ variation can also be seen in Table 5. Therefore, asteroseismology could help to probe the PP and CNO nuclear reactions in this stellar phase, leading us to have a better insight into the production of neutrinos.

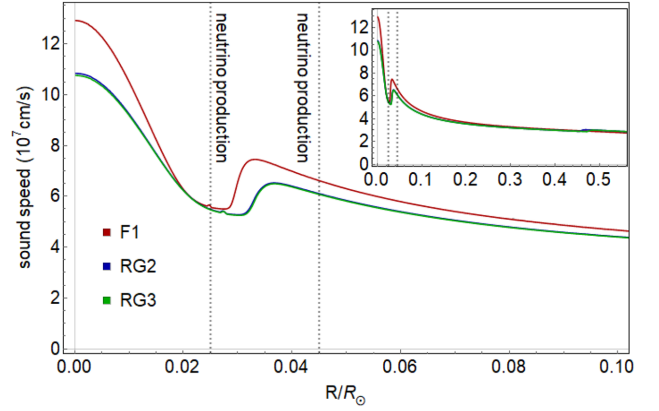

**Figure 3.** Absolute sound speed for all three models during the red giant phase for the extension of stellar radius where neutrino production is occurring. The inset shows a zoomed out version of the same plot.

For the three models described in the previous section, individual adiabatic oscillation modes were computed using GYRE (Townsend & Teitler 2013; Townsend, Goldstein & Zweibel 2018). When calculating $\Delta\nu$, the presented value is obtained from a linear fit to the frequencies as functions of radial order $n$, over the 22 radial orders around the frequency of maximum spectral intensity, $\nu_{\rm max}$. All quantities described above were obtained for the points indicated in Fig. 2 and can be found listed in Table 5. For all models, the relative standard deviation for $\Delta\nu$ as obtained from the linear fit of frequencies is always below 0.13 per cent for the MS points, and 0.5 per cent for the RG points.

Additionally, Fig. 4 shows the evolution of the (squared) Brunt–Väisälä frequency, showing the regions of active energy and neutrino production in the star, for ease of comparison with the results of Section 5. In both the main sequence (top panel) and red giant (bottom panel) phases, it can be seen that the regions where the relative differences between models are the most evident are the areas where nuclear reactions are occurring and element abundances are varying rapidly. Given the relation between the Brunt–Väisälä frequency and $\Delta\Pi$ presented in equation (4), the latter's success at discerning the various models is to be expected.

Data reported by Mosser et al. (2014) indicate that measurements of red giants can determine $\Delta\Pi_1$ with a precision between below 0.5 per cent for most cases and 2 per cent at worst, which would mean that the differences between the Low-Z and Low-Z' models would likely fall within the detection window, as they differ by 1.30 per cent, as would the difference between the High-Z and Low-Z models, numbering 2.58 per cent, assuming stars in the same conditions as those observed by the *Kepler* mission – and therefore, the precision expected to be obtained by the *PLATO* mission should be able to resolve this difference.

**Table 5.** Values of the large separation in frequency and period, and of relative neutrino luminosity for different models, for different stages of evolution.

| Model | | $\Delta\nu$ (μHz) | | | $\Delta\Pi_1$ (s) | | | $L_\nu/L$ (per cent) | | |
|---|---|---|---|---|---|---|---|---|---|---|
| | | Low-Z | High-Z | Low-Z' | Low-Z | High-Z | Low-Z' | Low-Z | High-Z | Low-Z' |
| 1 $M_\odot$ | MS | 136.07 | 137.77 | 132.59 | 1534.6 | 1513.6 | 1383.4 | 2.37 | 2.43 | 2.52 |
| | RG | 13.28 | 10.02 | 14.06 | 72.25 | 74.12 | 73.38 | 6.87 | 6.85 | 6.86 |
| | F1 | – | 5.847 | – | – | 62.09 | – | – | 6.84 | – |

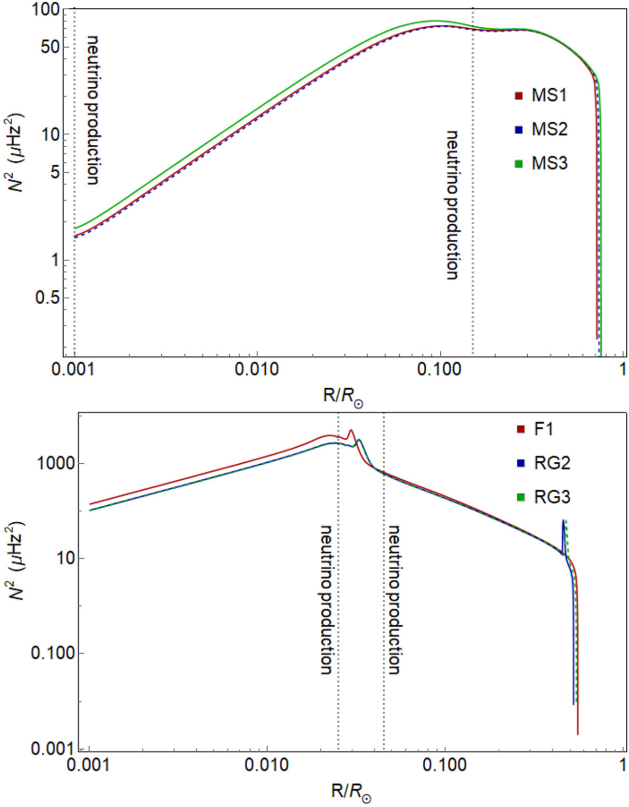

**Figure 4.** Square of the Brunt–Väisälä frequency for all three models, for the main sequence points (top) and for the red giant points (bottom). The region between the dotted vertical lines denotes the area of neutrino production.

**Table 6.** Predicted neutrino fluxes on the Earth by source of emission and comparison with other SSMs – V-H corresponds to B16-GS98 and V-L to B16-AGSS09met from Vinyoles et al. (2017) – and with observed values for the Sun from Bergström et al. (2016). Units are: $10^9$ ($^7$Be), $10^6$ ($^8$B, $^{17}$F), and $10^8$ ($^{13}$N, $^{15}$O) cm$^{-2}$ s$^{-1}$.

| Source | High-Z | V-H | Low-Z | V-L | Low-Z' | Sun |
|---|---|---|---|---|---|---|
| $\Phi(^7\text{Be})$ | 4.89 | 4.93 | 4.44 | 4.50 | 7.16 | $4.80 \pm 5$ per cent |
| $\Phi(^8\text{B})$ | 5.40 | 5.46 | 4.39 | 4.50 | 11.5 | $5.16 \pm 2.2$ per cent |
| $\Phi(^{13}\text{N})$ | 3.58 | 2.78 | 2.34 | 2.04 | 5.15 | <13.7 |
| $\Phi(^{15}\text{O})$ | 2.83 | 2.05 | 1.77 | 1.44 | 4.51 | <12.8 |
| $\Phi(^{17}\text{F})$ | 5.95 | 5.29 | 3.32 | 3.26 | 9.83 | <85 |

Comparison of the initial parameters for the models Low-Z and Low-Z' – especially $\alpha_i$ and $Y_i$ (cf. Table 3) – also reveals that these variables will have an important impact on the stellar surface (Lopes & Gough 1998, 2001; Brito & Lopes 2019), visible in the propagation of acoustic waves.

The similar values obtained for $L_\nu/L$ (cf. Table 5) can be explained by the fact that the criteria used to select the sets of points MS1, MS2, and MS3 or RG1, RG2, and RG3 were that of similar luminosity. As can be seen in Table 4, the corresponding models' other parameters (namely age and $T_{\text{eff}}$) may differ significantly, but in low-mass stars where neutrino production originates almost entirely from the nuclear processes (which is the case for the Sun), the total emission flux is strongly constrained by these processes, which also dictate the star's luminosity. One can argue that other criteria could have been chosen for the construction of the points to be compared but, at the later stages of evolution, no one parameter establishes a good enough description of the overall state of the star, and the use of collections of different parameters with different values quickly devolves to arbitrary fine-tuning. Therefore, we opted to follow an observational criterion by giving predominance to the luminosity.

## 5 NEUTRINOS AND THE SOLAR CORE

Since neutrino propagation in the Sun is unaffected directly by opacity or convection, study of neutrino spectra could provide a third way (in addition to electromagnetic spectroscopy and helio-seismology) to probe the abundances of the solar core bypassing the two largest currently hypothesized error factors, since carbon and nitrogen function as catalysts of the CNO cycle reactions, and so an increased abundance of these elements results in a higher emission of CNO neutrinos. Additionally, thanks to recent improvements in the measurements of the fluxes from $^7$Be and $^8$B decays, neutrinos provide an excellent test to the integrity of an SSM.

For those reasons, neutrino fluxes on the Earth for all three models were computed from the nuclear abundances and reaction rates provided by MESA. This is the first time this analysis is performed for MESA models, and these results are yet another indicator of the robustness of this stellar evolution code.

Solar observations of $^7$Be and $^8$B neutrino fluxes seem to indicate that their most likely value falls somewhere between the predictions of SSMs computed using the GS98 and the AGSS09 compositions and metallicites – a review of recent predictions can be found in Zhang, Li & Christensen-Dalsgaard (2019). Our results for High-Z and Low-Z models match these expectations as well, as can be seen in Table 6. These fluxes represent the total of neutrinos of all flavours, as they are usually presented in the literature, and comparison of these values with measurements of electronic neutrinos made on the Earth can be done by correcting these values with the parameters found in Bergström et al. (2016).

The relative fluxes for neutrinos by source are also presented, in Table 7, and show the increase of importance of the CNO neutrinos as the star evolves, due to the increase in relative importance of the CNO cycle for the star's energy production. The increase in temperature inside the star during this red giant phase leads to an increase of the reaction rates of the pp branch III reaction, and a decrease of the reaction rates of the pp branch II reaction. Consequently, this new arrangement leads to an increase of the $^8$B neutrino flux (produced by the pp branch III) and relative reduction of the $^7$Be one (produced by the pp branch II). A plot of the luminosity produced by pp and CNO reactions over time for all models can be seen in Fig. 5.

**Table 7.** Neutrino fluxes for the $^7$Be, $^8$B, $^{13}$N, $^{15}$O, and $^{17}$F sources relative to the respective predicted solar neutrino flux for the Low-Z model, for the present day, as seen in Table 6. The presented fluxes for the source $X_i$ for a model $\alpha$ at point $Y$ are in the form $\phi_{Y,\alpha}(X_i) = \Phi_{Y,\alpha}(X_i)/\Phi_{\text{MS, Low-Z}}(X_i)$, with $\Phi(X_i)$ being the neutrino flux from the source $X_i$.

| Model | | Age (Gyr) | | | $\phi(^7\text{Be})$ | | | $\phi(^8\text{B})$ | | |
|---|---|---|---|---|---|---|---|---|---|---|
| | | Low-Z | High-Z | Low-Z' | Low-Z | High-Z | Low-Z' | Low-Z | High-Z | Low-Z' |
| 1 $M_\odot$ | MS | 4.57 | 4.57 | 2.97 | 1.000 | 1.120 | 1.640 | 1.000 | 1.245 | 2.665 |
| | RG | 12.00 | 11.86 | 10.77 | 1.039 | 0.987 | 1.047 | 448.1 | 407.2 | 420.9 |
| | F | – | 12.00 | – | – | 0.424 | – | – | 629.9 | – |

| Model | | $\phi(^{13}\text{N})$ | | | $\phi(^{15}\text{O})$ | | | $\phi(^{17}\text{F})$ | | |
|---|---|---|---|---|---|---|---|---|---|---|
| | | Low-Z | High-Z | Low-Z' | Low-Z | High-Z | Low-Z' | Low-Z | High-Z | Low-Z' |
| 1 $M_\odot$ | MS | 1.000 | 1.525 | 2.195 | 1.000 | 1.587 | 2.529 | 1.000 | 1.791 | 2.959 |
| | RG | 1456 | 1463 | 1444 | 1923 | 1934 | 1903 | 446.0 | 484.0 | 513.8 |
| | F | – | 3708 | – | – | 4874 | – | – | 1115 | – |

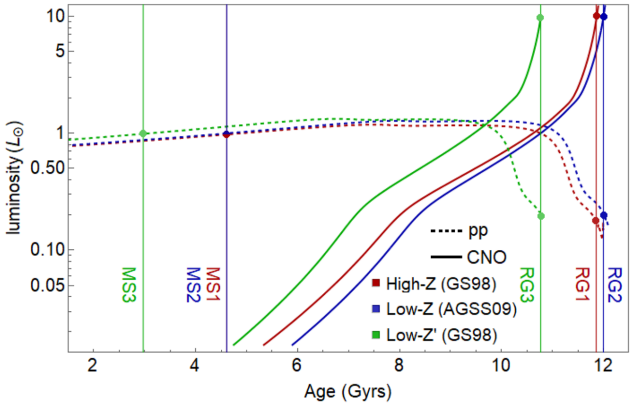

**Figure 5.** Luminosity (in units of $L_\odot$) of the pp and CNO reactions for the High-Z, Low-Z, and Low-Z' models. The increase in relative importance of the CNO cycle for the star's energy production can be seen towards the later ages.

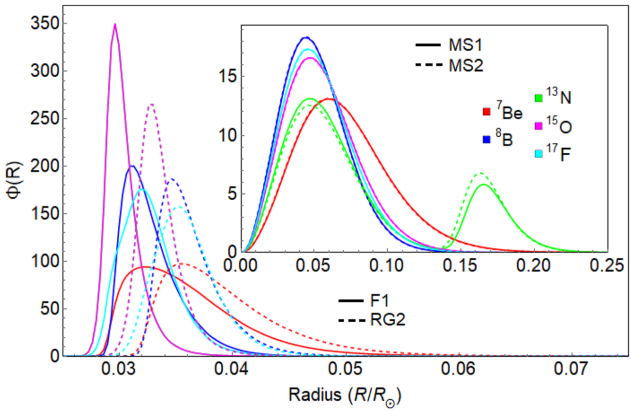

**Figure 6.** Radial neutrino flux comparison for the High-Z (full lines) and Low-Z (dashed lines) models for the indicated points: the inset for the current solar age (4.57 Gyr) and the outer plot for an age of 12 Gyr.

Figs 6 and 7 depict radial neutrino fluxes by source inside the Sun for the Low-Z and High-Z models, and for the Low-Z and Low-Z'

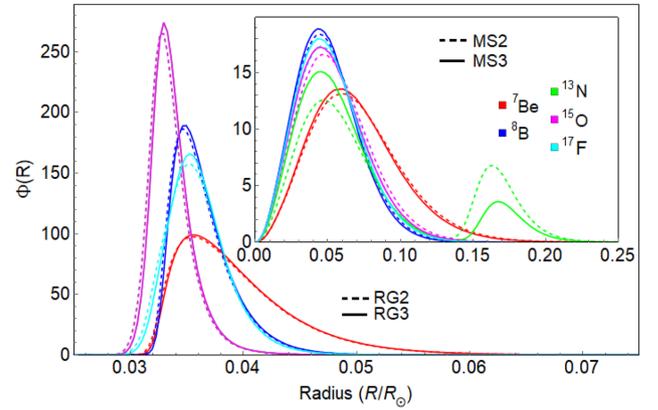

**Figure 7.** Radial neutrino flux comparison for the Low-Z' (full lines) and Low-Z (dashed lines) models for the indicated points: the inset for the point of current solar luminosity and the outer plot for a luminosity around 10.24 $L_\odot$.

models, respectively, for the indicated points from Fig. 2. The region where the peaks cluster in the outer (red giant) plot in Figs 6 and 7 corresponds to local maxima of the sound speed difference in the red giant phase, as can be qualitatively discerned from Fig. 3.

As expected, the possibility of observing the $^{13}$N, $^{15}$O, and $^{17}$F solar neutrinos proves to be outside the immediately accessible range of the current generation of neutrino experiments. However, advancements in the area of low-energy neutrino detectors and the expansion of current detector facilities could make it possible to reach this goal in the next 5–10 yr (Cerdeño et al. 2018). Lopes & Silk (2013) have also showed how different abundances of C, N, and O could lead to differences in the neutrino fluxes. Our predictions, regarding the CNO sector, place the fluxes of each relevant neutrino source for the High-Z and Low-Z models at a relative distance of 34.6, 37.5, and 45.2 per cent for $^{13}$N, $^{15}$O, and $^{17}$F, respectively, which means that a precision lower than 17.3 per cent in a future CNO-sensitive experiment would be able to select one of the two values while excluding the other, and provide the first good case for constraining the metallicity and composition choice based on neutrino observations.

## 6 CONCLUSIONS

Three models were built using the evolutionary code MESA to exemplify the impact of composition and metallicity choices at one point in the full evolution of the Sun: two SSMs, each following one of the solar composition determinations of Grevesse & Sauval (1998) (GS98) – High-*Z*, and Asplund et al. (2009) (AGSS09) – Low-*Z*, and one comparison model, using a metallicity of $Z = 0.0165$ but the relative element abundances described in GS98 – Low-*Z*', and calibrated to match the effective temperature and luminosity of the Low-*Z* model in the red giant stage. For these models, the surface and evolutionary parameters (age, *L*, *R*, and $T_{\rm eff}$), helioseismic indicators ($\Delta \nu$ and $\Delta \Pi_1$), and neutrino fluxes on the Earth were computed for different ages.

It has also been shown that stars of similar metallicities and exhibiting practically the same effective temperature and luminosity in the red giant stage can register a difference in age of up to 1 Gyr (or 8.3 per cent of stellar lifetime), with possible consequences to the dating of globular clusters' turn-off points that are used, among other things, to constrain the time-scales of galactic evolution and the age of the Universe.

From helioseismic analysis performed using GYRE, for stars in the red giant stage, we have shown that the current level of precision of asteroseismic measurements should be enough to differentiate all three models: even the Low-*Z* model from the Low-*Z*' model, which was purpose-built and calibrated to match its luminosity and effective temperature, differing only in the relative abundances of elements. Furthermore, a comparison of the sound speeds for both of these models in the region of nuclear activity (and therefore neutrino production) reveals a localized disagreement.

Moreover, we show that predictions for the neutrino flux on the Earth for the $^7$Be and $^8$B sources for our models follow the trend in the literature of overestimating (in case of the High-*Z* model) and underestimating (for the Low-*Z* model) the best estimations of current fluxes for those sources from observations. Our results also indicate that an eventual detection of the CNO neutrinos with a precision of around 17 per cent could be enough to solve the solar abundance problem, by singling out one of the model predictions, which are directly relatable to C and N content in the Sun.

**ACKNOWLEDGEMENTS**

The authors thank the Fundação para a Ciência e Tecnologia (FCT), Portugal, for the financial support to the Center for Astrophysics and Gravitation-CENTRA, Instituto Superior Técnico, Universidade de Lisboa, through the Grant Project No. UIDB/00099/2020.

They would like to acknowledge the efforts of Bill Paxton and the other 14 developers on the MESA team, and Rich Townsend, the developer of GYRE, for providing powerful software tools for free, for documenting, maintaining, and constantly improving the code, and for their availability to answer users' questions.

Additionally, thanks are owed to Ana Brito and José Lopes for invaluable discussions regarding solar modelling and calibration.

Finally, the authors would like to thank the anonymous referee for their detailed analysis and numerous suggestions on both the form and content of this work.

**DATA AVAILABILITY**

This article makes use of the MESA stellar evolutionary code that is available for free in the public domain and can be found at the Zenodo repository with the DOI: 10.5281/zenodo.3473377. All of the data sets used in this work can be found in the MESA files for the release version r12115. The GYRE code has also been used to compute oscillations, and is also available for free in the public domain at the GYRE Home and Wiki.

This paper has been typeset from a TeX/LaTeX file prepared by the author.